\documentclass[twocolumn,amsmath,aps,fleqn]{revtex4}

\usepackage{amssymb}
\usepackage{amsmath}
\usepackage[usenames]{color}
\usepackage{graphicx}

\usepackage{hyperref}

 \providecommand{\eprint}[1]{\href{http://arxiv.org/abs/#1}{#1}}
 \providecommand{\adsurl}[1]{\href{#1}{ADS}}
 
\providecommand{\url}[1]{\href{#1}{#1}}

\newcommand{\himpc}{\ensuremath{\,h\,{\rm Mpc}^{-1}}}

\providecommand{\eprint}[1]{\href{http://arxiv.org/abs/#1}{#1}}
 \providecommand{\adsurl}[1]{\href{#1}{ADS}}
 
\providecommand{\url}[1]{\href{#1}{#1}}

\newcommand\eq[1]{Eq.~(\ref{#1})}
\newcommand\eqs[2]{Eqs.~(\ref{#1}) and (\ref{#2})}

\newcommand{\beq}{\begin{equation}}
\newcommand{\eeq}{\end{equation}}
\newcommand{\bea}{\begin{eqnarray}}
\newcommand{\eea}{\end{eqnarray}}

\renewcommand{\d}{{\rm d}}

\newcommand{\mnras}{Mon. Not. R. Astron. Soc.}
\newcommand{\apjs}{Astrophys. J. S.}
\newcommand{\apjl}{Astrophys. J. Lett.}
\newcommand{\lsim}{\,\raise 0.4ex\hbox{$<$}\kern -0.8em\lower 0.62ex\hbox{$\sim$}\,}

\begin{document}
\title{Probing the Primordial Power Spectrum with Cluster Number Counts}
\author{Teeraparb Chantavat}
\affiliation{Oxford Astrophysics, Physics, DWB, Keble Road, Oxford, OX1 3RH, UK}
\author{Christopher Gordon}
\affiliation{Oxford Astrophysics, Physics, DWB, Keble Road, Oxford, OX1 3RH, UK}
\author{Joseph Silk}
\affiliation{Oxford Astrophysics, Physics, DWB, Keble Road, Oxford, OX1 3RH, UK}
	

\label{firstpage}

\begin{abstract}
We investigate how well galaxy cluster number counts can constrain the primordial power spectrum. 
Measurements of the primary anisotropies in the cosmic microwave background (CMB) may be limited, by the presence of foregrounds from secondary sources, to probing the primordial power spectrum at wave numbers less than about $0.30\, \himpc$.
 We break up the primordial power spectrum into a  number of nodes and interpolate linearly between each node. This allows us to show  that cluster number counts could then extend the  constraints on the form of  the  primordial power spectrum up to wave numbers of about $0.45\, \himpc$. 
We estimate combinations of constraints from PLANCK and SPT primary CMB and their respective SZ surveys. We find that their constraining ability is limited by uncertainties in the mass scaling relations. We also estimate the constraint from clusters detected from a SNAP like gravitational lensing survey. 
As there is an unambiguous and simple relationship between the filtered shear of the lensing survey and the cluster mass, 
it may be possible to obtain much tighter constraints on the primordial power spectrum in this case.
\end{abstract}

\maketitle
\section{Introduction}
A crucial element of cosmology is the form of the primordial fluctuations. These fluctuations provide the seeds for structure formation which we observe today through the cosmic microwave background (CMB), galaxy, and cluster surveys.
Current data is consistent with the primordial fluctuations being scalar, adiabatic, Gaussian and having a power 
spectrum with a simple power law parameterization \citep{spergel07,komatsu08}.
The primordial fluctuations may have been generated during a period known as `inflation', where the accelerated expansion
of the primordial Universe is driven by a potential dominated scalar field or fields (see for example \citet{lidlyt00}). If inflation was driven by a single scalar field with a  smooth potential, then the power spectrum of primordial fluctuations is predicted to be generally quite close to a power law form, although in some cases there may be significant running of the spectral index.
 However, if inflation was driven by multiple fields or by a single field with a feature in its potential, then the primordial power spectrum may  contain  hills, valleys, oscillations or other features (see for example \cite{salbonbar89,adarossar97,leach01,hunsar04,joy08,romsas08}).
 
The two main approaches to probing the primordial power spectrum are either to assume a specific form for a feature in the primordial power spectrum (see for example \cite{bond88,hamann07,hoiclihol07,hunsar07,joy08}) or to try and reconstruct the primordial power spectrum non-parametrically (see for example \cite{bridle03,huoka03,leach06,zhan06,spergel07,shasou08,verpei08}). The cleanest probe of the primordial power-spectrum is the cosmic microwave background (CMB). However, it is probably limited to  wave numbers smaller than about 0.30 $\himpc$ as beyond that  foreground contamination from secondary sources are likely to dominate the cosmological signal.

In this article, we investigate to what extent galaxy cluster number counts can probe the primordial power spectrum. Traditionally, the effect of the primordial power spectrum on clusters is probed with the constraints on $\sigma_8$, which is the dispersion of the linear theory matter fluctuations smoothed on scales of 8~$h^{-1}$Mpc. However, as we will discuss in Sec.~\ref{effectonn}, the value of $\sigma_8$ encompasses a very broad range of wave-numbers and so it is advantageous  to break up the primordial power spectrum into effectively several small bins. Additionally, $\sigma_8$ is sensitive to other cosmological parameters such as the matter density, the dark energy equation of state, primordial non-Gaussianity, and non-zero neutrino mass. Rather than simply comparing $\sigma_8$ inferred from clusters with that from inferred from the CMB, it may be better to introduce new parameters to account for the possible deviations from the fiducial model of $\Lambda$CDM consisting of a  featureless, adiabatic, and Gaussian primordial power spectrum. 
In this article we investigate how well the combination of the CMB and cluster surveys can constrain the primordial power spectrum to be featureless.

We begin in Section~\ref{sec:methods} with contrasting  how the cluster number counts and the primary CMB probe the primordial power spectrum. In Section~\ref{sec:fisher} we forecast the constraints on the primordial power spectrum from the Sunyaev-ZelÕdovich effect (SZ) PLANCK\footnote{http://www.rssd.esa.int/index.php?project=planck} and South Pole Telescope\footnote{http://pole.uchicago.edu/} (SPT) cluster surveys and combine them with the forecasted constraints on the primordial power spectrum from the PLANCK and SPT primary CMB survey. We also estimate the constraints from the  SNAP\footnote{http://snap.lbl.gov/}  lensing  cluster survey.
Concluding remarks are given in Section~\ref{sec:conclusions}.

\section{Dependence on Primordial Power Spectrum}
\label{sec:methods}
\begin{figure}
\includegraphics[width=7.7cm]{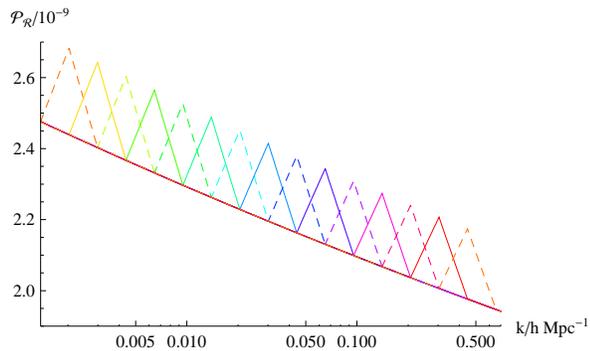}
\caption{\label{prim}The effect on the primordial power spectrum of individually changing  nodes 1 to 15 to a value of 1.1  in our feature function $F$. Each changed node is a plotted as a separate colour and odd numbered nodes are plotted as dashed lines.}
\end{figure}

The dimensionless primordial power spectrum, as function of the comoving wave number $k$, is usually parameterized as a power law
\beq
{\cal P}_{{\cal R},0}(k)\equiv{\Delta}_{{\cal R}}^{2}\left({k\over k_{\rm pivot}}\right)^{n_s-1}
\eeq 
where the amplitude (${\Delta}_{{\cal R}}^2$), spectral index ($n_s$), and pivot point ($ k_{\rm pivot}$) are taken to be independent of $k$.
We model deviations from this form in a similar way to \citet{bridle03} and \citet{spergel07}:
\beq
{\cal P}_{\cal R}(k)=F\left(k\right) {\cal P}_{{\cal R},0}(k)
\eeq
where the ``feature function'', $F(k)$, is specified by the values at 
17 nodes logarithmically spaced in $k$-space with 
\beq
{k_i \over h^{-1} {\rm Mpc}}={0.657 \over 1.47^i},\quad i=0,\ldots16\,.
\label{ki}
\eeq
Linear interpolation in $\log(k)$ is used to determine $F$ between each node. The effect of individually changing nodes 1 to 15 is shown in Fig.~\ref{prim}. Nodes 0 and 16 will always be fixed to 1.

\citet{spergel07}  (see their Fig.~11) found that WMAP III data constrained the nodes best at around $k\sim 0.01 \himpc$ with a 1 sigma error on $F$ of about 0.3. While at $k\sim 0.1 \himpc$ the  WMAP III data provided practically no constraint due to its relatively large beam size and measurement noise. It would be interesting to see how these constraints could be improved with the addition of small scale CMB experiments and other large scale structure data such as the Sloan Digital Sky Survey, but that is beyond the scope of the present paper. It is probably reasonable to say that $F$ is not known to better than 10\% precision with current data, certainly  for larger wave numbers $(k \gtrsim 0.1 \himpc)$ where the current constraints are likely to be significantly more uncertain. For this reason we take $F(k_i)=1.1$ to conservatively illustrate the effect of node changes in Figures \ref{prim}, \ref{dsig8}, \ref{dndz}, and \ref{dcmb}.

\subsection{Effect on number counts}
\label{effectonn}
The linear theory matter power spectrum at a redshift $z$ is given by
\beq
\label{Pm}
P(k,z)\propto T^2(k,z) k {\cal P}_{\cal R}(k)
\eeq
where $T(k,z)$ is the matter transfer function. 
We use CAMB\citep{lewchaslas00}\footnote{http://camb.info/} to evaluate $P(k,z)$ and we modified the ``ScalarPower'' function in CAMB to include our feature function $F(k)$. The accuracy and sample boost parameters in the CAMB initialization file where all set to the value of 3 which increases the precision and reduces the amount of interpolation. Although, there is  some interpolation in $k$  used by CAMB, the sampling is much smaller than the width  between our nodes. We include lensing of the CMB, but its effect is negligible for our primary region of interest, $\ell \leq 2000$.
Throughout this paper we assume a flat $\Lambda$CDM Universe with no tensor perturbations and we  use the WMAP5 maximum likelihood parameters \citep{dunkley08}: 
\bea
\Omega_b h^2=0.0227,\Omega_ch^2=0.108,n=0.961,\tau=0.089, \nonumber \\
\Delta_{\cal R}^2=2.41\times10^{-9},h=0.724,
\label{fid}
\eea
 where we have used the WMAP chosen pivot of $k_{\rm pivot}=(1/500) {\rm Mpc}^{-1}$ and the parameters have their usual meaning.

\begin{figure}
\includegraphics[width=8cm]{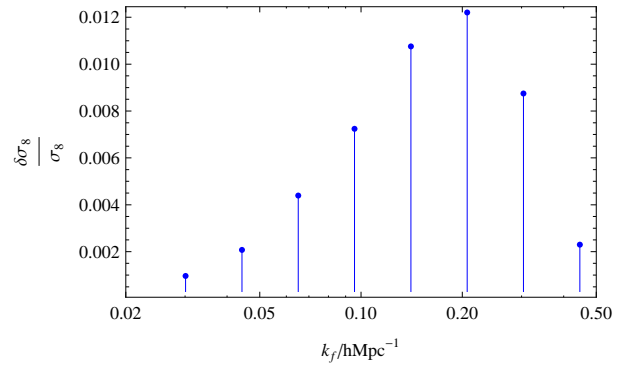}
\caption{\label{dsig8} Effect of individually setting the node at $F[k_f]=1.1$ on the variance of the matter field smoothed with $8\, h^{-1} {\rm Mpc}$ top hat window function.}
\end{figure}

The variance of the linear theory matter field, which has been smoothed by a top hat filter on a comoving length scale $R$, is given by
\beq
\sigma^2(R,z) = \int_{0}^{\infty}{{\rm d} k \over k}\, {k^{3} \over 2\pi^{2}}P(k,z)W^2(kR)
\label{sigma}
\eeq
where
\beq
W(kR)=3\left[  {\sin(kR) \over(kR)^{3}} -{\cos(kR)\over(kR)^{2}} \right] \, .
\label{tophat}
\eeq
The top hat smoothing suppresses the contribution of any fluctuations located at wave number $k_f\gg 1/R$.
Fig.~\ref{dsig8} illustrates how changing the primordial power spectrum alters $\sigma$ when $R=8h^{-1}$Mpc.
As can be seen, the main contribution is from the node at $k=0.20 \himpc$, but that there is also a reasonable contribution from several of the neighboring nodes. As will be discussed in Sec.~\ref{cmb}, the CMB is able to well constrain the primordial power spectrum at least for $k\leq 0.30\himpc$. So it follows that $\sigma_8$ is mainly affected by scales which are constrainable by the CMB. That is why our method of breaking up the primordial power spectrum into linearly interpolated nodes is useful as it gives enough flexibility to separate the parts of the primordial  power spectrum which affect cluster counts but may not be constrainable by the CMB. Also, $\sigma_8$ is sensitive to not only the primordial power spectrum but  also the other cosmological parameters, such as $\Omega_m=\Omega_c+\Omega_b$, the dark energy equation of state, primordial non-Gaussianity, and non-zero neutrino mass.

\begin{figure}
\includegraphics[width=8cm]{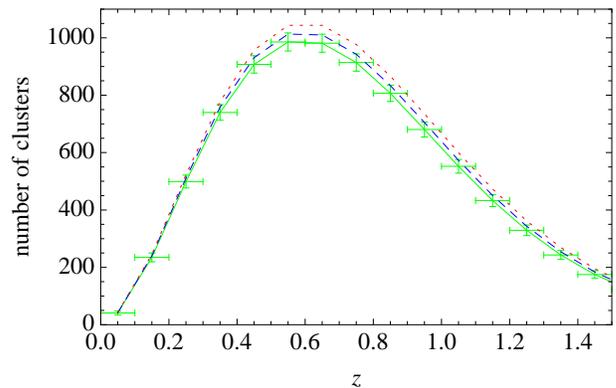}
\caption{\label{dndz} Effect of change in primordial power spectrum on the SNAP lensing cluster counts ($M_{\rm min}=10^{14}h^{-1} M_\odot$, $f_{\rm sky}=0.024$). The plots are for the featureless ($F=1$) power spectrum (green, solid), $F[0.45]=1.1$ (blue dashed), and $F[0.30]=1.1$ (red dotted). The vertical error bars are one sigma and the horizontal error bars indicate the bin width.}
\end{figure}

The number density of halos (bound objects) may be predicted using the smoothed linear theory density field  \citep{presch74}. For a background non-relativistic matter density of $\rho_m=\Omega_m \rho_{\rm total}$,
the number density ($n$) of halos of mass 
\begin{align}
M&={4\pi\over 3} R^3 \rho_m\nonumber \\&=1.16\times10^{12}\Omega_{m}h^{-1}\left({R\over h^{-1}Ê{\rm Mpc}}  \right)^{3}M_{\odot}
\label{massscale}
\end{align}
 depends, to a good approximation, on the primordial power spectrum only through its effect on $\sigma(R,z)$
  \citep{presch74,shetor99,shemotor99,kneislsil01,jenkins01}
\beq
{\d n(z)\over \d M} = {\rho_m \over M} {\d\ln\sigma(z)^{-1}\over \d M} f(\sigma(z))\, .
\eeq
 We use the Sheth-Tormen mass function
  \citep{shetor99,shemotor99} 
  for which
\beq
f=A\sqrt{2a \over \pi}\left[ 1+\left({\sigma(z)^{2}\over a\delta_{c}^{2}} \right)^{p} \right]{\delta_{c} \over \sigma(z)}\exp\left(-{a\delta_{c}^{2}\over 2\sigma(z)^{2}}\right)
\label{fst}
\eeq
where $A=0.3222,a=0.707,p=0.3,$ and $\delta_{c}=1.686$.
The top hat smoothing in \eq{sigma} suppresses the contribution of any change to the primordial power spectrum located at wave number $k_f\gg 1/R$. Combined with \eq{massscale}, this implies that a change in the primordial power spectrum at $k_f$ has a suppressed effect on the number density on mass scales satisfying
\beq
{ M \over h^{-1}M_{\odot}  }\gg 10^{12}  \left ({k_{f} \over h {\rm Mpc}^{-1}}
\right)^{-3} \, .
\label{maxmass}
\eeq
The number of clusters per redshift interval above some mass threshold $M_{\rm min}$ is given by
\beq
\frac{dN}{dz}(M>M_{\rm min})= f_{\rm sky} \frac{dV(z)}{dz}\int_{M_{\rm min}}^\infty\! dM \frac{dn}{dM}(M,z)\, .
\label{dNdz}
\eeq
Where $f_{\rm sky}$ is the fraction of the sky being observed and the volume element is given by
\beq
\frac{dV}{dz}=\frac{4\pi}{H(z)}\left[\int_0^z \frac{dz'}{ H(z')}\right]^2
\eeq
and $H(z)$ is the Hubble parameter
\begin{equation}
  H(z) = H_0\sqrt{\left( {\Omega_{m}(1+z)^3 + ( 1 - \Omega_{m})} \right)}\, .
\end{equation}
The effect of a change in the primordial power spectrum on the number counts is illustrated in Fig.~\ref{dndz} for a SNAP-like gravitational lensing
cluster survey. 
The SNAP selection function can be approximated by setting $M_{\rm min}=10^{14}h^{-1} M_\odot$ in \eq{dNdz} (see Section~\ref{sec:fisher}). Due to the exponential suppression in \eq{fst}, $\d N/\d z$ mainly depends on the mass scale $M_{\rm min}$. From \eq{maxmass} this implies the SNAP cluster survey will become insensitive to the primordial power spectrum for $k\gg0.2\himpc$. This is consistent with $F[0.45 ]=1.1$ having less of an effect than  $F[0.30]=1.1$ as illustrated in Fig.~\ref{dndz}.

\subsection{Effect on the CMB}
\label{cmb}
\begin{figure}
\includegraphics[width=8cm]{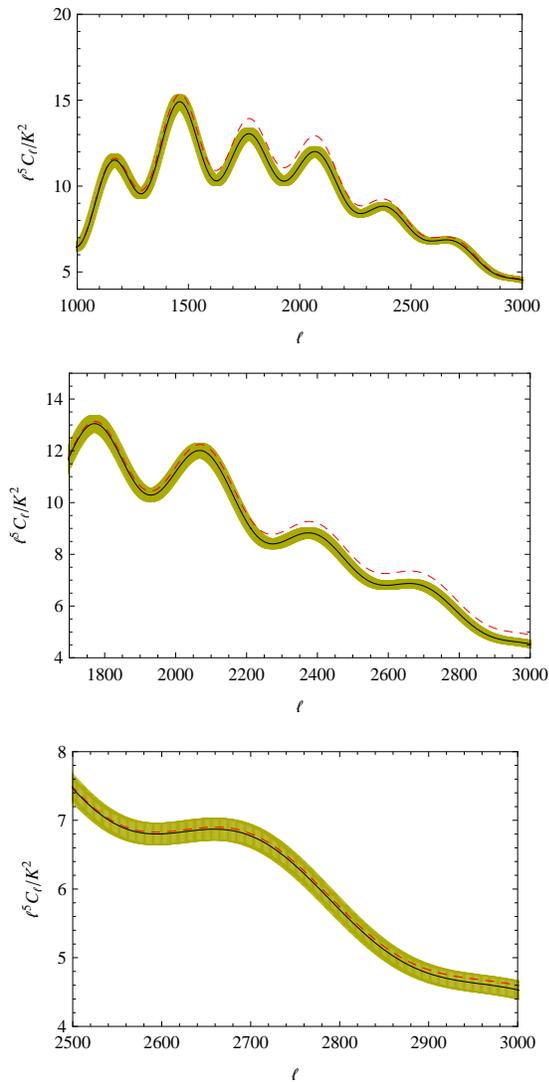}
\caption{\label{dcmb} Effect of a change in primordial power spectrum on the CMB (TT) angular power spectrum. Error bars (yellow) include instrument noise, beam size, and cosmic variance effects. They are centered on the  featureless ($F=1$) (black) power spectrum and are for SPT. The red dashed curves are for
 $F[0.21]=1.1$ (top panel), $F[0.30]=1.1$ (middle panel), and $F[0.45]=1.1$ (bottom panel). Perfect subtraction of secondary foreground sources is assumed for the figure}
\end{figure}

The  primordial power spectrum is probed over a wide range of wave numbers by measurements of the primary CMB anisotropies  (see for example \citet{huoka03}). Both the temperature ($T$) and E-mode of the polarization ($E$) probe scalar perturbations. 
\beq
{\ell(\ell+1)C_{\ell}^{XX'} \over 2\pi} = \int \d \ln k\, T^X_\ell(k)T^{X'}_\ell(k) {\cal P}_{\cal R}(k)
\eeq
where $X,X'\in T,E$.
The projection of a mode of wave-number $k$ on to the surface of last scattering  (a sphere of comoving radius $D_*$) results in the CMB transfer functions having the form $T^X_\ell \sim j_\ell(kD_*)$. Where $j_\ell$ is the spherical Bessel function of order $\ell$ which peaks at $\ell\approx kD_*$. Therefore a feature in the primordial power spectrum at wave-number $k_f$ is mapped onto a feature in CMB angular power spectrum at
\beq 
\ell\sim k_fD_*\approx 10^4 {k_f \over  h {\rm Mpc}^{-1}}
\label{cmbl}
\eeq 
We also used our modified version of CAMB  to evaluate $C_{\ell}^{XX'} $. Although, there is some interpolation in $\ell$  used by CAMB, the sampling is much smaller than the width between our nodes. Fig.~\ref{dcmb} shows the effect $F\not=1$ on $C_{\ell}^{TT}$ with predicted SPT error bars (see Section~\ref{sec:fisher}). 
The panels in the figure are consistent with \eqs{ki}{cmbl} in that having $F(k_i)=1.1$ translates roughly into a triangular perturbation of about 10\% amplitude in $C_{\ell}$.
In Fig.~\ref{dcmb} we have not included the foreground contribution from secondary sources which will probably be hard to completely remove for $\ell>2000$. For this reason, as done by \citet{huoka03} and \citet{leach06}, we 
will restrict ourselves to $\ell\leq2000$ when evaluating the forecasted marginalized errors.

\section{Forecasts}
\label{sec:fisher}
We use the Fisher matrix formalism 
 to make forecasts on how well the primordial power spectrum can be constrained. 
We take as our fiducial model  the WMAP5 maximum likelihood parameters, \eq{fid}. 
We consider two cluster count SZ experiments: PLANCK and SPT.  For PLANCK, the selection function can be approximated by 
$f_{\rm sky}=0.8$ and $M_{\rm min, PLANCK}=5\times 10^{14} h^{-1} M_\odot$ \citep{geihob06}. 
A common approximation for the SPT selection function is 
 $f_{\rm sky}=0.1$ and $M_{\rm min, SPT}=1.75\times 10^{14} h^{-1} M_\odot$
(see for example \citet{sefusatti07,loverde08}). 
This may be overly optimistic, but increasing it to $M_{\rm min, SPT}=3\times 10^{14} h^{-1} M_\odot$ does not qualitatively change our conclusions as the uncertainty in the mass scaling relationship turns out to be the limiting factor.  
 We take  $z\in [0,3]$ with bin sizes  $\delta z=0.1$, although the bulk of the constraining power comes from $z<1$
and the results are negligibly changed if we take $z<2$. We also forecast the SNAP cluster lensing survey  constraints \citep{marber06} (also see \citet{hamtakyos04,wang04,fanhai07,takbri07}). Here we take $z<1.5$, $f_{\rm sky}=0.024$, and $M_{\rm min, SNAP}=10^{14} h^{-1} M_\odot$. This roughly matches the 
number of clusters found using a more accurate selection function of \citet{marber06} when we use their fiducial model cosmological parameters. For our fiducial  model, \eq{fid}, we predict a total of
8888 clusters detected from a SNAP cluster survey.  Also, \cite{takbri07} found that the signal to noise ratio of a more realistic selection function was about the same as taking a mass limit of $10^{14} h^{-1} M_\odot$.
Our SNAP selection function is biased to slightly higher redshifts than that of \citet{marber06}, but we expect this not to alter our predicted constraints significantly.

The number of clusters  in redshift bin $i$ is Poisson distributed with the expected number ($e_i$) given by integrating \eq{dNdz} over the red shift bin. 
  Element $(j,k)$  of the Fisher matrix for a cluster count experiment is given by \citep{holhaimoh01}
\beq
{\cal F}_{jk}=\sum_{i=1}^{N_{\rm bins}} {1\over e_i} {\partial e_i \over \partial p_j}{\partial e_i \over \partial p_k}
\eeq
where  $p_j$ consists of the cosmological parameters in \eq{fid} except for $\Delta_{\cal R}^2$ and $n$ as they will be almost completely degenerate with the feature function $F$.
 To account for uncertainties in the mass of the SZ observed clusters, we also allow $M_{\rm min,PLANCK}$ and $M_{\rm min,SPT}$ to be free parameters and give them both priors of 10\% one sigma errors. 
 The lensing observed clusters have a well determined mass-scaling relation and so we do not take $M_{\rm min,SNAP}$  to be a free parameter \citep{hamtakyos04,wang04,fanhai07,takbri07}.
 Additionally, we allow the nodes 1 to 15 in \eq{ki} of $F$ to vary. The derivatives are taking at the fiducial values of the parameters which in the case of the $M_{\rm min}$ are the previously specified values and for the feature function, $F(k_i)=1$ for all $i$. The derivatives are approximated by the symmetrized form of a difference equation so as to minimize truncation error (see for example \citet{presteuvet92}).
 

The CMB  Fisher matrix is given by (see for example \citep{zalspesel97})
\beq
{\cal F}_{ij}=\sum_{\ell} \sum_{X,X'} {\partial C_{\ell}^X \over \partial p_i} {\rm Cov}^{-1} (C_{\ell}^X,C_{\ell}^{X'}){\partial C_{\ell}^{X'} \over \partial p_j }
\eeq
where the covariance matrix can be obtained from \citet{zalspesel97} and it depends on the temperature  noise per pixel ($\sigma_T$), the polarization noise  per pixel ($\sigma_E$), 
the pixel area in radians squared ($\theta^2=4\pi/N_{\rm pix}$),
and the beam window function which we approximate as Gaussian ($B_\ell\approx \exp(-\ell(\ell+1)\sigma_b^2$). The values we use are taken from the PLANCK blue book\footnote{http://www.rssd.esa.int/SA/PLANCK/docs/Bluebook-ESA-SCI(2005)1\_V2.pdf} and are listed in Table~\ref{planckparam} (note that $\theta$ needs to be converted to radians).
\begin{table}\centering
\caption{\label{planckparam} PLANCK Instrument Characteristics \label{planckparm}}
\begin{tabular}{lcccc}
\hline
Center Frequency (GHz) &70&100&143&217\\ \hline
$\theta$ (FWHM arcmin) &14&10&7.1&5.0\\
$\sigma_T$ ($\mu{\rm K}$)& 12.8& 6.8& 6.0& 13.1\\
$\sigma_E$ ($\mu{\rm K}$)&18.2&10.9&11.4& 26.7 \\
\hline
\end{tabular}
\end{table}
We use $\sigma_b=\theta/\sqrt{8 \log[2]}$ and combine the different frequency bands as specified in \citet{bonefsteg97}.  We also include the constraints from 
SPT primary CMB temperature measurement for which we just use one band with $\theta=1$~arcmin and $\sigma_T=10\mu{\rm K}$. We take the range in $\ell$ to be 2 to 2000. At higher $\ell$, secondary sources of temperature and polarization will likely prohibit the extraction of cosmological information from the primary CMB. 

\begin{figure}
\includegraphics[width=8cm]{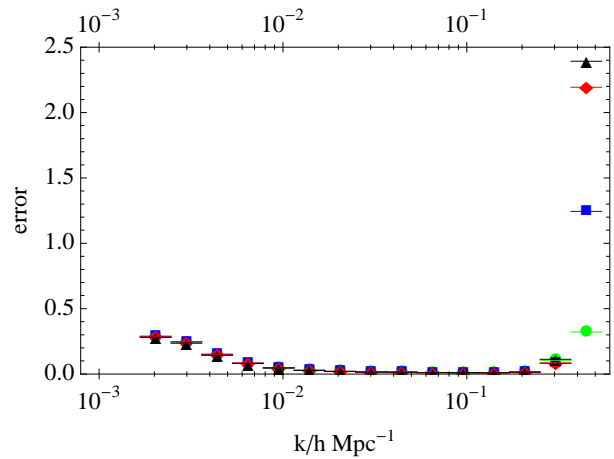}
\caption{\label{results} Expected marginalized one sigma errors for the different nodes of feature function ($F$). Errors are shown for PLANCK (TT, TE, EE) (black triangles), PLANCK (TT, TE, EE) and SPT (TT) (red diamonds), PLANCK (TT,TE,EE, clusters) and SPT (TT, clusters) (blue squares), and PLANCK (TT, TE, EE) and   SNAP (clusters) (green circles).}
\end{figure}

The  expected covariance matrix of the parameter errors is approximated by the inverse of the Fisher matrix. The expected marginalized one sigma error bars are then given by the square roots of the diagonal elements of the expected covariance matrix. Also, experiments can be combined by adding the Fisher matrices. When we combine PLANCK and SPT, primary CMB or SZ cluster detection, we reduce the $f_{\rm sky}$ for PLANCK to 0.7 so as not to count the same clusters twice. We plot the expected one sigma marginalized errors for each node of $F$ in Fig.~\ref{results}. For each node, marginalization is done over all other nodes, cosmological and mass parameters.
The lack of constraint at low $k$ is from cosmic variance. At high $k$, the primary CMB does not constrain the primordial power spectrum as it is assumed to be limited to $\ell\leq 2000$ due to secondary sources. As shown in Fig.~\ref{results}, including the  lensing detected clusters makes a big improvement in constraining the primordial power spectrum at $k=0.45 h{\rm Mpc}^{-1}$. The PLANCK (TT, TE, EE) data only constrains
the primordial power spectrum at $k=0.45 h{\rm Mpc}^{-1}$ to about $250\%$. 
The constraints on the  other cosmological parameters we included are not significantly altered by the addition of clusters. Combining all the surveys, we considered, improves 
the constraint at $k=0.45 h{\rm Mpc}^{-1}$ to about $25\%$.

\begin{figure}
\includegraphics[width=7.5cm]{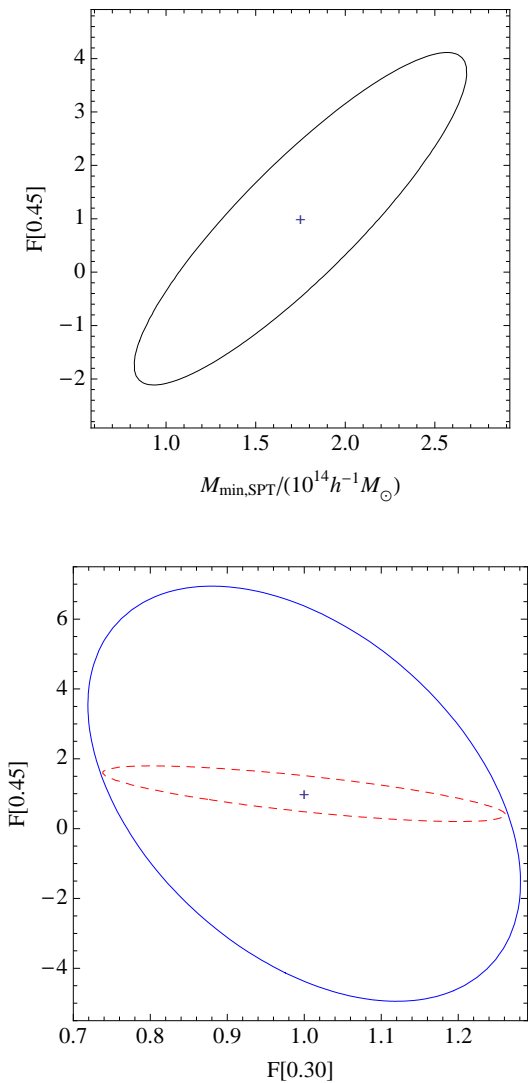}
\caption{\label{results2d}Marginalized probability contours containing 95\% of the expected posterior probability. 
The top panel is for PLANCK (TT, TE, EE) and SPT (clusters). The bottom panel is for  PLANCK (TT, TE, EE) alone (blue solid) and 
with SNAP (clusters) (red dashed)).
}
\end{figure}

The main reason why the SZ cluster surveys are not as effective at constraining the primordial power spectrum as the cluster lensing survey is due to a degeneracy with the uncertainty in their mass parameters, see  the top panel of Fig.~\ref{results2d}. As the primary CMB is taken to be limited to $\ell \leq 2000$, it is barely altered by changes to the node at $F(0.45)$, (see Fig.~\ref{dcmb}).
As can be seen from the bottom panel of Fig.~\ref{results2d}, the addition of the SNAP lensing cluster survey dramatically sharpens the constraint at $F(0.45)$. This is consistent with the large effect seen in Fig.~\ref{dndz}.
The primary CMB would need to be removed of foregrounds to a high level of accuracy up to $\ell\sim 3000$ in order to accurately measure $F(0.45)$ without the aid of clusters (see Fig.~\ref{dcmb}).

\section{Conclusions}
\label{sec:conclusions}
In this article we have investigated what role cluster number counts can play in constraining the primordial power spectrum. 
We found that if PLANCK and SPT primary CMB anisotropy measurements are limited to $\ell \leq 2000$ by secondary source foregrounds, then they can only measure the primordial power spectrum 
at $k=0.45\, h{\rm Mpc}^{-1}$  to about 220\% precision due to the degeneracy between changes in the primordial power spectrum at smaller $k$.
Including SPT and PLANCK SZ cluster surveys increases the precision to about $124\%$, but they are limited by a degeneracy with the determination of the observed clusters' masses. While a SNAP like gravitational lensing cluster survey combined with PLANCK primary CMB data may be able to increase the precision to about 30\% due to the accurate relationship between the observed lensing shear and cluster masses.

In the current article we have used  simple minimum mass selection functions for the cluster surveys. As we are investigating a new use for cluster surveys, we think it is justifiable to initially get a more qualitative and easily reproducible forecasted constraint. In future work, building on the current investigation,
 we will 
evaluate the effect on our current conclusions of more realistic selection functions  such as those in \citep{pierre08,marber06}.


It is common in cluster surveys or forecasts of surveys  to evaluate the constraints on $\sigma_8$ with and without the CMB. If there is a feature at high $k$ in the primordial power spectrum then this could lead to a discrepancy in the value of $\sigma_8$ obtained from the CMB. Our method could then be used to determine the size of the feature needed to explain such a discrepancy. Other possible sources in a discrepancy between the CMB and cluster constraints on $\sigma_8$ could be a non-cosmological constant source of dark energy,  primordial non-Gausianity, and sufficiently large neutrino mass. Degeneracies between the dark energy equation of state and non-Gaussianity where looked at by \citet{sefusatti07}. They found that, provided redshift information was available, there was not significant degeneracy between the two.
Our method could  be useful in determining what the observational degeneracies between features and other possible  sources of discrepancy in $\sigma_8$ are. Comparing our Fig.~\ref{dndz} with Fig.~2 of \citet{sefusatti07} indicates that there may be some degeneracy between a feature in the primordial power spectrum and primordial non-Gaussianity. Also, inflation models which generate features in the primordial power spectrum may naturally generate scale dependent non-Guassianity \citep{cheeaslim08}. This could increase the overall change in cluster counts and thus make the feature more easily detectable.

An alternative way of parameterizing features in the primordial power spectrum is to allow a running of the spectral index, $\d n_s /\d \ln k$ (see for example \citep{lidlyt00}). This is less flexible than our current approach, but may be more natural to implement in an inflation model. We plan to investigate how well cluster number counts improve the running of the spectral index in future work.

{\em Acknowledgements:}
We thank Dick Bond, Jo Dunkley,  Sam Leach, Lance Miller, Florian Pacaud, Anze Slosar, Wiley Wang, and Joe Zuntz for helpful discussions. TC is funded by The Institute for the Promotion of Teaching and Science and Technology (IPST) in Thailand. CG is funded by the Beecroft Institute for
Particle Astrophysics and Cosmology.


\end{document}